# A Survey on Factors Affecting Iran's Fuel Rationing Smart Card User Acceptance and Security

Seyed Gholamreza Eslami
Centre for Advanced Software Engineering
Universiti Teknologi Malaysia
54100 Kuala Lumpur, Malaysia
reza9750@yahoo.com

Ali Peiravi
Department of Electrical Engineering
Ferdowsi University of Mashhad
Mashhad, Iran
Peiravi@ferdowsi.um.ac.ir

Behzad Molavi
Department of Electrical Engineering
Imam Reza University of Mashhad
Mashhad, Iran
Behzad_molavi2006@yahoo.com

*Abstract-* **Smart card technology has resulted in vast developments in many aspects of modern human life. User acceptance of fuel rationing smart cards based on adoption model involves many factors such as: satisfaction, security, external variables, attitude toward using, etc. In this study, user acceptance and security factors for fuel rationing smart cards in Iran have been evaluated based on an adoption model by distributing a questionnaire among UTM (University Technology Malaysia) Iranian students, MMU (Multimedia University) Iranian students, either asking by e-mail from people who are not available.**

*Keywords-smart card, fuel rationing, user acceptance, security, satisfaction, external varaibles, attitude toward use, adoption model*

I.

## I. INTRODUCTION

Fuel subsidies are common in many countries especially in the Middle East rich oil nations. The rapid growth in population, the increased number of cars, the extended use of private vehicles for transportation, the increased smuggling of fuel due to price difference in Iran and neighboring countries, etc. led to the dire need to establish a fuel subsidy reduction plan in Iran using a fuel rationing smart card system.

The introduction of smart card technology for fuel rationing in Iran dates back to 2007 when the government and the parliament decided to implement the first phase of a plan for the reduction of subsidies. This type of fuel rationing smart card is to be used by car owners at the gas station. The plan has led to certain advantages such as reduction of total fuel distribution by managing vehicle owners' fuel quota and limiting smuggling of fuel to neighboring countries.

A smart card is a plastic card, its size is the same as a credit card (smart card dimension is 85.6 mm by 54 mm) but it also has a memory chip embedded inside it for reading and writing. Smart cards have capabilities like storing and saving data, making calculations, processing data, managing files, and executing encryption algorithms. With the improvement in smart card technology, smart cards will be replacing cash, identification cards, visas and passports, airline tickets, and medical records for patients (Al-Alawi and Al-Amer, 2006). Smart cards have been used for fuel distribution system to define each vehicle's quota. They also enable proper supervision and control of the fuel distribution system since all transactions with the fuel stations are saved. By using fuel rationing smart cards, we can check the amount of fuel that each driver has received. When a driver inserts the fuel rationing smart card into the smart card reader of the gas pump, the balance of his fuel quota will be shown. The gas distribution authorities can also monitor all fuel transactions.

The issue of user acceptance of new technologies by end users is a major issue in actual implementation of products of modern technology. Arami et al. (2004) studied the user acceptance of multifunctional cards amongst the students of the Vienna University of Economics and Business Administration (WU Vienna). Many researchers have focused on the development of models for new technology adoption in various parts of the world. For example, Wahed (2007) used the technology adoption model to analyze internet adoption and use among men and women in Indonesia. Although the fuel rationing plan in Iran may not be the best as argued by Rahdari and others (2009) who consider this to be only a symptomatic solution, issues such as user acceptance and security of the implemented plan are important since this plan has been implemented. Mohammadi (2009) developed an adoption model to assess smart card technology acceptance. Yeow and Loo (2009) studied the factors affecting user acceptance of smart cards in Malaysia with the objective of identifying user acceptance factors influencing Malaysians' intention to use MyKad IC. Their intent was to present recommendations to the government to increase acceptance. Holden (2010) discussed technology acceptance model in health care to evaluate end user's interest in health information technology. Security is an integral part of user acceptance factor. Vijayasarathy (2004) argued that security is one of the most significant factors for technology acceptance. Thus, in this study we have analyzed the factors that affect security of fuel rationing smart cards separately.

## II. FUEL RATIONING SMART CARD TECHNOLOGY

A fuel rationing smart card is a plastic card, just at the size of a credit card (Effing, 2003), with a memory chip embedded inside it. The chip holds data with appropriate security. This data is associated with either value or information or both and is stored and processed within the fuel rationing smart card's chip.

A fuel rationing smart card should have capabilities such as saving and retrieving data. We need to save all fuel system transaction on memory of fuel rationing smart card due to fuel transaction tracking. We need to retrieve data from memory of fuel rationing smart card to check the correctness of the amount of fuel that a driver has filled to the tank. For this purpose we need to have a microchip on fuel rationing smart card.

The implementation of the fuel rationing system in Iran started in Zahedan, i.e. a province in the south east of Iran. Then it was extended to the whole country in a short period of time. Based on statistics, on Jan. 17th, 2007 around 8000 fuel pumps in 2300 fuel stations were equipped with smart card readers. A fuel smart card imposes the fuel quota set up by the government for each vehicle. This quota is based on the type of vehicle and its intended use. For instance, the amount of fuel quota for private vehicles is different from that of public vehicles such as taxis, ambulances, and buses that receive a lot more fuel.

Upon inserting a fuel rationing smart card into the card reader unit at the pump, the fuel balance is displayed for the user. The system is able to provide users fuel at three different prices based on the quota predetermined by the government.

## III. RESEARCH METHODOLOGY

In this study, a survey was conducted to analyze the security and user acceptance of fuel rationing smart card technology as adopted in Iran. The questionnaire consists of seventy-three items in two sections. It was distributed among the forty-one Iranian students of Centre for Advanced Software Engineering (CASE) at University Technology of Malaysia (UTM) and Multimedia University (MMU) in the field of computer science, software engineering and MBA. Questionnaires were also sent out by an e-mail to nine unavailable respondents who were Iranian workers in ASHNA Company and N.I.O.P.D.C.

The first section of the designed questionnaire includes seven questions for assessing demographic characteristics such as age, gender, level of education, category of fuel rationing smart card user, and card user's experience in working with the fuel rationing smart card. Section two of the questionnaire includes sixty-six questions about factors that affect fuel rationing smart card user acceptance and security with the format of; strongly disagree, disagree, neither agree nor disagree, agree, and strongly agree, so 3 is the neutral point. Respondents should choose the answer based on their personal opinion.

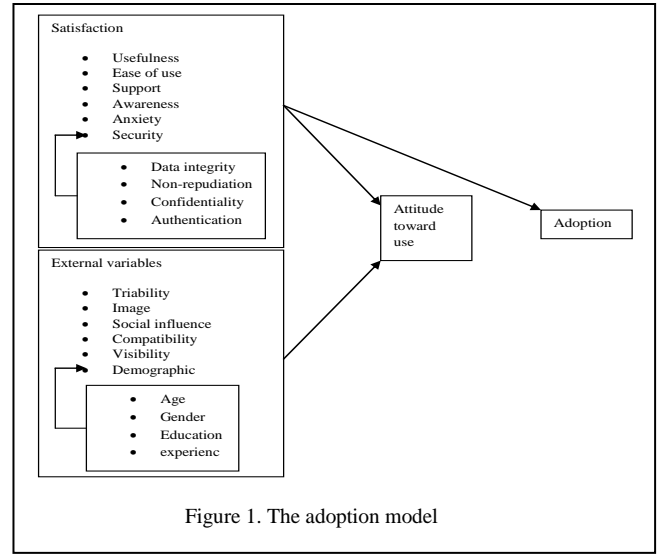

Figure 1. The adoption model

## IV. CONCEPTUAL FRAMEWORK

The research model developed by Mohammadi (2009) shown in Fig. 1 was adopted as the conceptual framework for this study. This conceptual framework is based on the technology acceptance model (TAM), Innovation Diffusion Theory (IDT), Unified Theory of Acceptance and Use of Technology (UTAUT), Extension of Technology Acceptance Model (TAM2), and Theory of Reasoned Action (TRA) (Taherdust, 2009). Table 1 shows the involved variables and their sources.

Table 1. Variables and their sources.

| Factors | Definition | Source(s) |
|---|---|---|
| Awareness | The degree to which an individual aware about the technology. | Al-Alawi and Al-Amer Bandura |
| Support | The degree to which an individual believes that an organizational and technical infrastructure exist to support use of the system. | Bailey and Pearson Al-Gahtani.et.al |
| Anxiety | The degree to which users are worried about using technology. | Bailey and Pearson Igbaria.et.al |
| Ease of use | The degree to which a person believes that using a particular system is free of effort. | Davis Bailey and Pearson |

| Usefulness | The degree to which a person believes that using a particular system would enhance his or her job performance. | Davis<br>Bailey and Pearson<br>Venkatesh and Davis |
|---|---|---|
| Security | The degree to which a person feels that security is important to them and believes that using smart card is secure. | Bailey and Pearson<br>Vijayasarathy |
| Compatibility | The degree to which the innovation is perceived to be consistent with the potential user's existing values, previous experience and needs. | Rogers |
| Image | The degree to which use of an innovation is perceived to enhance one's image or status in one's social system. | Moore and Benbasat<br>Venkatesh and Davis |
| Social influence | The degree to which an individual perceives that it is important to other believes he or she use the new system. | Ajzen<br>Venkatesh and Davis |
| Triability | The degree to which an innovation may be experimented before adoption. | Rogers |
| Visibility | The degree to which the results of an innovation are visible and communicable to others. | Rogers<br>Moore and Benbasat |
| Demographic | Age, Gender, Education, Experience. | Agarwal and Prasad<br>Venkatesh and Davis<br>Taylor and Todd |
| Confidentiality | Ensuring that data are only accessible to those authorized to receive them. | Hendry |
| Authentication | The process of specifying identity of person. | Hendry |
| Data integrity | Confirms that correctness of message transmitted from the origination to the destination. | Hendry |
| Non-repudiation | Confirms that the origin of data is exchanged in transaction. | Hendry |
| Attitude toward behavior | An individual negative or positive feeling about execute the target behavior. | Fishbein and Ajzen |
| Adoption | User's adoption of information technology is dependent on their perceived ease of use and perceived usefulness of the technology. | Mohammadi |

## V. RESULTS AND DISCUSSION

Table 2 summarizes the demographic profile and descriptive statistics of the respondents. It shows that 36% of the respondents are male and 64% of them are female.

Gender differs in the relative influence of attitude towards technology use, subjective norm (social influence) and perceived behavioral control (Venkatesh and Davis, 2000); decisions of men were strongly affected by attitude towards using technology and women's decisions were more strongly influenced by subjective norms and perceived behavioral control. Table 2 shows that 88% of the respondents are not more than 35 years old. Computer skills were more easily learned by younger person compare to older persons (Czara et. Al. 1989). So working with fuel rationing smart card is more acceptable by younger people. 86% of the respondents have at least a bachelor's degree. By increasing smart card users' level of education their tendency for use of smart cards is increased (Raub, 1981; Igbaria and parasuraman, 1989; Haward, 1988). 82% of the respondents are students who are considered to be the informed part of each society. All respondents have used fuel rationing smart cards before and 82% of them have used it for at least one year. 68% of the respondents have used fuel rationing smart card for at least once a week.

Table 3 shows the mean and the P-value for all factors and their sub-factors. Based on this table, security is the most important sub-factor for fuel rationing smart card users. In addition, authentication is the most important sub-factor among the security factors. All user acceptance sub-factors and security sub-factors are significant based on the respondents' opinion. External variables and satisfaction sub-factors are not significant based on the respondents' opinion.

Based on Table 4, security is the most important factor for respondents aged between 31-35, female and respondents with a PhD degree. Based on Table 5, user acceptance is the most important factor for respondents aged over 35, male and respondents with a PhD degree.

Table 2. Demographic profile of respondents

| Demographic variables | Frequency (N) | Percentage (%) |
|---|---|---|
| **Gender** | | |
| Male | 18 | 36 |
| Female | 32 | 64 |
| **Age** | | |
| 20 or under | 9 | 18 |
| 21-25 | 9 | 18 |
| 26-30 | 11 | 22 |
| 31-35 | 15 | 30 |
| More than 35 | 6 | 12 |
| **Education** | | |
| Diploma degree | 7 | 14 |
| Bachelor Degree | 20 | 40 |
| Master Degree | 19 | 38 |
| PhD Degree | 4 | 8 |
| **Category** | | |
| UTM Iranian Student | 20 | 40 |
| MMU Iranian Student | 21 | 42 |
| Resident in Mashhad | 6 | 12 |
| ASHNA Employee | 2 | 4 |
| N.I.O.P.D.C Employee | 1 | 2 |
| **How familiar are you with fuel rationing smart card?** | | |
| Never heard about it. | 0 | 0 |
| I have heard but never used it. | 0 | 0 |
| Use it only sometimes. | 35 | 70 |
| Use it on a regular basis. | 15 | 30 |
| **How long have you been using fuel rationing smart card?** | | |
| Less than a month | 3 | 6 |
| More than a month | 6 | 12 |
| About a year | 23 | 46 |
| More than a year | 18 | 36 |
| **How frequent do you use fuel rationing smart card?** | | |
| More than once a day | 5 | 10 |
| Once a day | 8 | 16 |
| Weekly | 21 | 42 |
| Monthly | 16 | 32 |

Table 3. Mean and P-value for factors and their sub-factors

| Factors and their sub factor(s) | Mean | Std. Deviation | Sig. (2-tailed) |
|---|---|---|---|
| **Satisfaction** | 4.0572 | 0.16086 | 0.000 |
| Usefulness | 4.11 | 0.2306 | 0.182 |
| Ease of use | 4.11 | 0.31875 | |
| Support | 4.0633 | 0.34975 | |
| Awareness | 3.955 | 0.46206 | |
| Anxiety | 3.985 | 0.44438 | |
| **Security** | 4.3329 | 0.23537 | 0.000 |
| Authentication | 4.58 | 0.38279 | |
| Confidentiality | 4.2267 | 0.44385 | |
| Non-repudiation | 4.15 | 0.63286 | |
| Data integrity | 4.28 | 0.4779 | |
| **External variables** | 3.9973 | 0.24719 | 0.000 |
| Trial ability | 3.9533 | 0.64597 | 0.015 |
| Image | 4.1467 | 0.45256 | |
| Social influence | 4.1067 | 0.53639 | |
| Compatibility | 3.97 | 0.43342 | |
| Visibility | 3.81 | 0.57045 | |
| **Attitude toward use** | 3.9160 | 0.37491 | 0.000 |
| **Adoption** | 2.93 | 0.68146 | 0.000 |
| **User acceptance** | 3.847 | 0.18170 | 0.000 |

Table 4. Mean and P-value for security based on demographic factors

| Demographic factors | | Mean | Std. Deviation | Sig. (2-tailed) |
|---|---|---|---|---|
| Age | 20 or under | 4.0763 | 0.349 | 0.000 |
| | 21-25 | 4.2437 | 0.16313 | |
| | 26-30 | 4.2994 | 0.12721 | |
| | 31-35 | 4.5093 | 0.8318 | |
| | More than 35 | 4.4722 | 0.414 | |
| Gender | Male | 4.2170 | 0.301 | 0.058 |
| | Female | 4.3981 | 0.1603 | |
| Education | Diploma | 4.2933 | 0.1246 | 0.023 |
| | Bachelor | 4.2277 | 0.3102 | |
| | Master | 4.4214 | 0.12582 | |
| | PhD | 4.5083 | 0.03543 | |

Table 5. Mean and P-value for user acceptance based on demographic factors

| Demographic factors | | Mean | Std. Deviation | Sig. (2-tailed) |
|---|---|---|---|---|
| Age | 20 or under | 3.7342 | 0.1849 | 0.124 |
| | 21-25 | 3.8258 | 0.18807 | |
| | 26-30 | 3.8644 | 0.1552 | |
| | 31-35 | 3.8591 | 0.16038 | |
| | More than 35 | 3.9833 | 0.21202 | |
| Gender | Male | 3.8885 | 0.1746 | 0.917 |
| | Female | 3.8232 | 0.18407 | |
| Education | Diploma | 3.8592 | 0.14826 | 0.535 |
| | Bachelor | 3.8021 | 0.19675 | |
| | Master | 3.8746 | 0.16147 | |
| | PhD | 3.9151 | 0.26251 | |

## VI. Conclusion

In this study, we analyzed the factors affecting user acceptance and security for fuel rationing smart cards in Iran. Based on the adoption model, acceptance factor consists of five factors and security factor consists of four factors. We have used one-way ANOVA test in our assessment. Based on the results obtained from the analysis of the filled questionnaires, the respondents place more importance on the security of fuel rationing smart cards compared to other factors. In addition, since user acceptance of fuel rationing smart cards is very important, we have to pay attention to user acceptance factor of fuel smart card. In the future, more work should be done to expand the number of people surveyed and use a broader category of respondents. In addition, the passage of time since the initial implementation of this plan has affected the people's opinions and this effect should also be studied. Moreover, the government in Iran is about to implement the second phase of the fuel subsidy reduction plan and its effects should also be studied.